\documentclass[final,5p,twocolumn]{elsarticle}

\usepackage{lineno,hyperref}
\usepackage{amsmath,amsfonts,amsthm}

\newcommand{\bRp}{\mathbb{R}_+}

\newcommand{\bRN}{\mathbb{R}^3}
\newcommand{\Dint}{\int_{\mathbb{R}^6 \times \mathbb{R}_+^2 \times [0,1]^2 \times S^{2}}}
\newcommand{\dsvi}{ \mathrm{d}r \, \mathrm{d}R \, \mathrm{d}I_*  \, \mathrm{d}I  \, \mathrm{d}\mathbf{v}_*\,  \mathrm{d}\mathbf{v}}

\newcommand{\beqs}{\begin{equation*}}
\newcommand{\eeqs}{\end{equation*}}
\newcommand{\beq}{\begin{equation}}
\newcommand{\eeq}{\end{equation}}
\newcommand{\bx}{\mathbf{x}}
\newcommand{\bv}{\mathbf{v}}
\newcommand{\bc}{\mathbf{c}}
\newcommand{\bu}{\mathbf{u}}
\newcommand{\bg}{\mathbf{g}}

\newcommand{\bG}{\mathbf{G}}

\newcommand{\bU}{\mathbf{U}}
\newcommand{\bom}{\boldsymbol{\omega}}
\newcommand{\bsig}{\boldsymbol{\sigma}}
\newcommand{\md}{\mathrm{d}}
\newcommand{\lp}{\left(}
\newcommand{\rp}{\right)}
\newcommand{\la}{\left|}
\newcommand{\ra}{\right|}

\newcommand{\sjNi}{\sum_{\begin{subarray}{l} j=1\\j\neq i\end{subarray}}^{I} }

\newcommand{\korenij}{\sqrt{\frac{2 \, R  E}{\mu_{ij}}}}

\newcommand{\inzv}{\left|\,\mathbf{v} \right|}

\newcommand{\inzvp}{\left| \mathbf{v}' \right|}

\newcommand{\inzvui}{\left| \mathbf{v} - \mathbf{u}_i \right|}

\newcommand{\inzvsuj}{\left| \mathbf{v}_* - \mathbf{u}_j \right|}
\newcommand{\inzvMvs}{\left| \mathbf{v} - \mathbf{v}_* \right|}

\newcommand{\inzgMuiMuj}{\left| \mathbf{g} - \lp \mathbf{u}_i - \mathbf{u}_j \rp \right|}
\newcommand{\inzuiMuj}{\left| \mathbf{u}_i - \mathbf{u}_j \right|}
\newcommand{\inzg}{\left| \mathbf{g} \right|}

\newcommand{\Mi}{\mathbf{N}_i}

\newcommand{\Nij}{{\mathbf{N}}_{ij}}
\newcommand{\cNij}{\Omega_{ij}}
\newcommand{\Ei}{E_i}

\newcommand{\aij}{{\alpha}_{ij}}
\newcommand{\bij}{{\beta}_{ij}}

\newcommand{\sums}{\sum_{\begin{subarray}{l} j=1\\j\neq i\end{subarray}}^{I} }

\newcommand{\sumsl}{\sum_{\begin{subarray}{l} \ell=1\\\ell\neq i\end{subarray}}^{I} }

\newcommand{\sumskm}{\sum_{\begin{subarray}{l} k=1\\k\neq i\end{subarray}}^{I-1} }

\newcommand{\cjed}{K_{ij}^1}
\newcommand{\cdva}{K_{ij}^2}
\newcommand{\ctri}{K_{ij}^3}

\newcommand{\const}{C_{ij}}

\newcommand{\constil}{C_{i\ell}}

\begin{document}

\begin{frontmatter}

\title{Multi-velocity and multi-temperature model of the mixture of polyatomic gases issuing from  kinetic theory}

\author{Milana Pavi\'c-\v Coli\'c\corref{mycorrespondingauthor}}
\address{Department of Mathematics and Informatics \\ Faculty of Sciences, University of Novi Sad \\ Trg Dositeja Obradovi\'ca 4, 21000 Novi Sad, Serbia}
\cortext[mycorrespondingauthor]{Corresponding author}
\ead{milana.pavic@dmi.uns.ac.rs}


\begin{abstract}
In this paper, we consider Euler-like balance laws for mixture components that involve macroscopic velocities and temperatures for each different species. These laws are not conservative due to mutual interaction between species. In particular, source terms that describe the rate of change of momentum and energy of the constituents appear. These source terms are computed with the help of kinetic theory for mixtures of polyatomic gases. Moreover, if we restrict the attention to processes which occur in the neighborhood of the average velocity and temperature of the mixture, the phenomenological coefficients of extended thermodynamics can be determined from the computed source terms.
\end{abstract}

\begin{keyword}
mixtures, polyatomic gases, kinetic theory, source terms, phenomenological coefficients
\end{keyword}

\end{frontmatter}


\section{Introduction}

We consider a mixture of $I$ polyatomic rarefied gases, denoted with $\mathcal{A}_i$, $i=1,\dots,I$. 
We  study the behavior of its components, since it is known that  the behavior of a mixture as a whole at the macroscopic level can be very different from the behavior of its components when they are observed separately. 

Within the continuum theories, the most sensitive question is about velocity and temperature field variables. In that sense, we distinguish between \textit{i) single-temperature approach} \cite{groot-book}, that stays within the framework of classical thermodynamics and assumes one macroscopic temperature of mixture and one macroscopic velocity and  therefore, the state of the mixture is determined by the mass densities $\rho_i$ of each constituent, the mixture velocity $\bu$ and the common temperature $T$, and \textit{ii) multi-temperature approach}  in the context of rational extended thermodynamics that  introduces temperature $T_i$ and velocity $\bu_i$ for each component of the mixture \cite{ruggMTsurvey,rugg-sug-book}.


Multi-velocity and multi-temperature models issuing from extended thermodynamics \cite{rugg-ss-07,ruggMTsurvey}, obey three principles proposed by Truesdell in \cite{trusdel69}, namely  (i) all properties of the mixture must be mathematical consequences of properties of the constituents, (ii) to describe the motion of a constituent, we may in imagination isolate it from the rest of the mixture, provided that we properly take into account for the actions of the other constituents upon it, (iii) the motion of the mixture is governed by the same equations as is a single body. 


In particular, the second principle implies that each component of the mixture obeys balance laws of mass, momentum and energy, that are not conservative, because of the mutual interaction of the constituents. Thus, when chemical reactions are excluded from  consideration,  the following laws hold
\begin{equation}\label{polyT: balance laws constituents}
\small
\begin{split}
\partial_t  \rho_i  + \nabla_\bx \cdot \left( \rho_i   \bu_i \right) &= 0, \\
\partial_t \left(  \rho_i \bu_i \right) + \nabla_\bx \left( \rho_i \bu_i \otimes \bu_i +  \mathbf{p}_i  \right) & = \Mi, \\
\partial_t \rho_i\left(\tfrac{\left|\bu_i\right|^2 }{2} + e_{i} \right) + \nabla_\bx \cdot \left\{ \left(\rho_{i}\left(\tfrac{\left|\bu_i\right|^2}{2}    + e_{i} \right) +\mathbf{p}_{i} \right) \bu_{i}  + \mathbf{q}_i  \right\} & = E_i,
\end{split}
\end{equation}
for  $i=1,\dots,I$, with 
$\rho_i$ being the density, $\bu_i$ the velocity, $e_i$ the internal energy, $\mathbf{p}_i$ the  pressure tensor and $\mathbf{q}_i$ the heat flux vector of the species $\mathcal{A}_i$. The source terms $\Mi$ and $\Ei$ correspond to momentum, respectively energy exchange. 

In this paper we consider Eulerian fluids when off-diagonal parts of the pressure tensor and the heat flux vector vanish,
\beq\label{polyT: ass Eulerian}
\mathbf{p}_i = p_i \mathbf{I}, \qquad  \mathbf{q}_i=0, \quad  \,i =1,\dots,I,
\eeq
where $p_i$ is the hydrodynamic pressure of the constituent $\mathcal{A}_i$, and $\mathbf{I}$ is the identity matrix.\\

The source terms $\Mi$ and $E_i$  need to satisfy the following relations
\begin{equation}\label{polyT: sum (production terms) = 0}
\sum_{i=1}^I \Mi = \mathbf{0}, \qquad \sum_{i=1}^I \Ei=0,
\end{equation}
since the first Truesdell's principle asserts that the whole is just a sum of its parts, which means that we need to recover conservation laws of mass, momentum and energy for the mixture as a whole by summing equations \eqref{polyT: balance laws constituents}. This is precisely achieved by imposing the  restriction \eqref{polyT: sum (production terms) = 0} and by defining 
\begin{itemize}
	\item mass density of mixture $\displaystyle \rho =\sum_{i=I}^I \rho_i $,
	\item mixture velocity $\displaystyle \bu  = \frac{1}{\rho} \sum_{i=1}^I \rho_i \bu_i$, 
	\item diffusion velocity $\bU_i  = \bu_i - \bu$, with $\displaystyle \sum_{i=1}^I \rho_i \bU_i =0$,
	\item {pressure tensor} $\displaystyle \mathbf{p}  =\sum_{i=1}^I \left( \mathbf{p}_i  + \rho_i \bU_i \otimes \bU_i \right)$, 
	\item internal energy $ \displaystyle e  = \frac{1}{\rho} \sum_{i=1}^s \rho_i \left( e_i + \tfrac{1}{2} \left|\bU_i\right|^2 \right)$,
	\item flux of internal energy\\ $ \displaystyle \mathbf{q} =\sum_{i=1}^I  \left( \left( \rho_i  e_i + \frac{\rho_i}{2}  \left|\bU_i\right|^2  + \mathbf{p}_i \right) \bU_i + \mathbf{q}_i \right).$
\end{itemize}
Thus, summation of \eqref{polyT: balance laws constituents} over $i=1,\dots,I$, yields conservation laws for the mixture 
\begin{equation}\label{polyT: BL mixture u opstem obliku}
\begin{split}
\partial_t  \rho  + \nabla_\bx \cdot \left( \rho  \bu \right) &= 0, \\
\partial_t \left( \rho \bu \right) + \nabla_\bx \left( \rho \bu \otimes \bu + \mathbf{p} \right) & = 0, \\
\partial_t \left(\tfrac{\rho}{2}  \left|\bu\right|^2 + \rho e \right) + \nabla_\bx \cdot \left\{\left(\tfrac{\rho}{2}   \left|\bu\right|^2 + \rho e \right) \bu + \mathbf{p} \bu  + \mathbf{q} \right\} & = 0,
\end{split}
\end{equation}
that are identical to those
of a single fluid,   according to the third principle.\\

With the assumption  \eqref{polyT: ass Eulerian}, the system \eqref{polyT: balance laws constituents} is still not closed, since we need to determine the source terms $\Mi$ and $E_i$. 

One approach to the closure problem is issuing from extended thermodynamics \cite{rugg-ss-07}, where the objectivity and entropy principle are exploited in order to obtain their structure. The  model consists of balance laws \eqref{polyT: balance laws constituents} for species $\mathcal{A}_i$, $i=1,\dots,I-1$ and mixture conservation laws \eqref{polyT: BL mixture u opstem obliku} that replaces balance law for species $\mathcal{A}_I$, together with the assumption \eqref{polyT: ass Eulerian} of Eulerian fluids. 
The source terms obtained in \cite{rugg-ss-07} are
\beq\label{polyT: production terms ET}
\begin{split}
	\Mi &= - \sum_{j=1}^{I-1} \aij(\boldsymbol{w}) \lp \frac{\bu_j}{T_j} - \frac{\bu_I}{T_I} -\bu \lp \frac{1}{T_j} - \frac{1}{T_I} \rp \rp,\\
	\quad \Ei &=\bu \cdot \Mi - \sum_{j=1}^{I-1} \bij(\boldsymbol{w}) \lp - \frac{1}{T_j} + \frac{1}{T_I} \rp,  
\end{split}
\eeq
for all $ i=1,\dots,I-1$, where $T_i$ is temperature of species $\mathcal{A}_i$ (connected with pressure via $p_i=n_i k T_i$, $n_i$ is the number density and $k$ is the Boltzmann constant), $\boldsymbol\alpha=\left[\alpha_{ij}\right]_{1\leq i,j \leq I-1}$ and $\boldsymbol\beta=\left[\beta_{ij}\right]_{1\leq i,j \leq I-1}$ are positive definite $(I-1)\times(I-1)$ matrix functions of the objective quantities $\boldsymbol{w}$ (i.e. quantities  invariant with respect to the Euclidean transformations). Thus,  the given  model contains phenomenological coefficients $\boldsymbol\alpha$ and $\boldsymbol\beta$, and  extended thermodynamics does not provide any mean for their explicit determination. 

On the other side, the source terms $\Mi$ and $E_i$ can be determined using the kinetic theory of gases, provided that the collisional cross-section is specified. In particular, in  \cite{Bm1, Bm2} these source terms are calculated for some specific choices of the cross-section, and polyatomic gases are modeled with discrete energy levels.\\

The goal of this paper is to calculate the production terms $\Mi$ and $E_i$ starting  from the continuous internal energy model in the kinetic theory of gases from \cite{des-mon-sal}. Furthermore, the determined source terms are compared to the ones in  \eqref{polyT: production terms ET} in order to derive explicit formula for phenomenological coefficients $\boldsymbol\alpha$ and $\boldsymbol\beta$ evaluated at the local equilibrium state.  \\

The relation between extended thermodynamics of polyatomic gases and kinetic theory was analyzed in \cite{ATRS-Meixner, PRS, rugg-sug-book}. So far, the analysis of phenomenological coefficients was successfully solved only at the level of binary mixture \cite{damir1}.\\

The plan of the paper is as follows. We present kinetic model in the Section \ref{Sec: Kin model}, that allow us to compute the source terms in the Section \ref{Sec: compute}. These source terms are compared to the ones coming out from extended thermodynamics in \ref{Sec: comparison} in a linearized setting. In particular, we obtain phenomenological coefficients of thermodynamic model.

\section{Kinetic Model  for mixtures of polyatomic gases with continuous internal energy}\label{Sec: Kin model}

In the kinetic theory, the state of mixture component  $\mathcal{A}_i$ is described by a distribution function $f_i\geq0$, $i=1,\dots,I$. In this paper, we follow the model with continuous internal energy  presented in  \cite{des-mon-sal}. Thus, the distribution function in this case,  $f_i:=f_i(t,\bx,\bv,I)$,   depends on time $t\geq0$, space position $\bx \in \mathbb{R}^3$, velocity $\bv \in \mathbb{R}^3$ and the so-called  microscopic internal energy $I\geq0$, that aims at capturing all phenomena related to the polyatomic gas features (for example, rotations or vibrations during collision process).

In the kinetic theory style, the distribution function $f_i$ changes due to the binary collisions with other particles of species $\mathcal{A}_j$, $j\in \left\{ 1, \dots, I\right\}$. As a measure of its change, multi-species  collision operators are introduced. Therefore, the evolution of the distribution function is governed by the Boltzmann equation
\begin{equation}\label{BE_i}
\partial_t f_i + v \cdot \nabla_x f_i = \sum_{j=1}^I Q_{ij}(f_i, f_j)(\bv, I), \quad 1\leq i \leq I,
\end{equation}
where $Q_{ij}(f_i, f_j)$ is the collision operator that describes interaction of molecules of species $\mathcal{A}_i$ with molecules of species $\mathcal{A}_j$ described by distribution function $f_j$. It reads  
\begin{multline}\label{collision operator}
Q_{ij} (f_i, f_j) (\bv, I) =  \int_{\mathbb{R}^3 \times \mathbb{R}_+ \times [0,1]^2 \times S^{2}} \left[ f'_i f'_{j*} - f_i  f_{j*} \right] \\ \times \mathcal{B}_{ij} \,  \left( 1-R \right) R^{\frac{1}{2}}  \frac{1}{\varphi_i (I)} \,\mathrm{d} \omega  \mathrm{d}r \mathrm{d} R \mathrm{d}I_* \mathrm{d}\bv_*,
\end{multline}
with standard abbreviations $f_{j*}:= f_j(t,x,\bv_*,I_*)$,  $f'_{i}:= f_i(t,x,\bv',I')$, $f'_{j*}:= f_j(t,x,\bv'_*,I'_*)$, where $\bv',  \bv'_*, I', I'_*$ are given in terms of $\bv, \bv_*, I, I_*$ and parameters $\bom \in S^2$, $r, R \in [0,1]$ via collisional rules
\begin{equation}\label{polyT: full collision transformation}
\begin{split}
\bv' &= \bG +  \frac{m_j}{m_i + m_j}  \sqrt{\frac{2 \, R  E}{\mu_{ij}}} \, T_{{\omega}} \hat{\bg}, 
\\ \bv'_* &= \bG - \frac{m_i}{m_i + m_j} \sqrt{\frac{2 \, R  E}{\mu_{ij}}} \, T_{{\omega}} \hat{\bg}, \\
I'&=r(1-R)E, \qquad
I'_* = (1-r) (1-R)E, 
\end{split}
\end{equation}
with the total energy (kinetic plus microscopic internal energy) of the pair of particles during a collision
\begin{equation*}
E:= \frac{\mu_{ij}}{2} \left|\bg\right|^2  + I + I_* = \frac{\mu_{ij}}{2}   \left|\bg'\right|^2  + I' + I'_*,
\end{equation*}
and 
where $\bG$ is a vector of center of mass $ \bG:=\frac{m_i \bv+ m_j \bv_*}{m_i + m_j}$, $\bg$ the relative velocity $\bg:= \bv-\bv_*$, and $\hat{\phantom{a}}$ denotes the unit vector i.e. $\hat{\bg}:= \bg/\left|\bg\right|$, $\mu_{ij}$ is the reduced mass $\mu_{ij}:= \frac{m_i m_j}{m_i + m_j}$, and the mapping $T_{\omega} \mathbf{z} = \mathbf{z} - 2 \left( \bom \cdot \mathbf{z} \right) \bom $, $\forall \, \mathbf{z} \in \mathbb{R}^3$. The cross section $\mathcal{B}_{ij}:=\mathcal{B}_{ij} \left( \bv, \bv_* , I, I_*, r, R, \bom \right)$ is supposed to satisfy the micro-reversibility assumptions:
\begin{equation}\label{cross section micro assump}
\begin{split}
	\mathcal{B}_{ij} \left( \bv, \bv_* , I, I_*, r, R, \bom \right) &= \mathcal{B}_{ji} \left( \bv_*, \bv, I_*, I, 1-r, R, \bom \right),\\
	 &= \mathcal{B}_{ij} \left( \bv', \bv'_*, I', I'_*, r', R', \bom \right).
\end{split}
\end{equation}
The test function $\varphi_{i}(I)$ will be chosen in order to recover perfect gas law for polyatomic gases in equilibrium.
\subsection{Weak form of collision operator}
Taking the moment of the collision operator \eqref{collision operator} weighted with $\varphi_{i}(I)$ against some test function $\psi_i(\bv,I)$ yields
\begin{multline}\label{weak form}
\int_{\bRN \times \bRp} Q_{ij} (f_i, f_j) (\bv, I) \, \psi_i(\bv,I) \, \varphi_{i}(I)  \, \md I \, \md \bv 
\\=  \int_{\mathbb{R}^6 \times \mathbb{R}_+^2 \times [0,1]^2 \times S^{2}} f_i  f_{j*} \left( \psi_i(\bv',I') - \psi_i(\bv,I) \right) \\ \times \mathcal{B}_{ij} \,  \left( 1-R \right) R^{\frac{1}{2}}   \,\mathrm{d} \omega  \mathrm{d}r \mathrm{d} R \mathrm{d}I_* \md I \mathrm{d}\bv_* \md \bv.
\end{multline}

\subsection{Macroscopic quantities and conservation laws for mixtures}
In order to recover macroscopic balance laws for mixture components \eqref{polyT: balance laws constituents} from the kinetic theory point of view, we first define macroscopic quantities. Mass density, momentum density and total energy density of the species $\mathcal{A}_i$ are defined as the following moments of the distribution function:
\begin{equation}\label{polyT: macro eq}
\left(  \begin{matrix}
	\rho_i\\
	\rho_i \bu_i\\
	\frac{\rho_i}{2}  \left|\bu_i\right|^2 + \rho_i e_i
\end{matrix} \right)
= \int_{\mathbb{R}^3 \times \mathbb{R}_+}\left(\begin{matrix}
	m_i \\ m_i \bv \\ \frac{m_i}{2} \left|\bv\right|^2 +I
\end{matrix}\right)  f_i \, \varphi_i(I) \, \mathrm{d} I \, \mathrm{d} \bv.
\end{equation}
If we introduce the peculiar velocity that corresponds to the specie $\mathcal{A}_i$ with $\bc_i=\bv - \bu_i$, then pressure tensor and heat flux corresponding to the species $\mathcal{A}_i$ are  defined as follows
\begin{equation}\label{polyT: macro eq press and heat}
\left( \begin{matrix}
\left[\mathbf{p}_i\right]_{k\ell}\\
\mathbf{q}_i
\end{matrix}\right)
=  \int_{\mathbb{R}^3 \times \mathbb{R}_+}  \left( \begin{matrix}
m_i \, \left[\bc_i\right]_k \left[\bc_i\right]_{\ell} \\
\left( \tfrac{m_i}{2} \left|\bc_i\right|^2 + I \right) \bc_i
\end{matrix} \right) f_i \, \varphi_i(I) \, \mathrm{d} I \, \mathrm{d} \bv.
\end{equation}
Now, the macroscopic balance laws \eqref{polyT: balance laws constituents} can be obtained from the Boltzmann equation  \eqref{BE_i}  in the following way: we integrate it with respect to $\bv \in \mathbb{R}^3$ and $I \in \mathbb{R}_+$, previously multiplying  it  with the test function $\varphi_i(I)$ and with (i) $m_i$  to obtain  \eqref{polyT: balance laws constituents}$_1$, (ii) $m_i \bv$   to get  \eqref{polyT: balance laws constituents}$_2$, and (iii) $(\frac{m_i}{2} \left|\bv\right|^2 +I)$  to obtain  \eqref{polyT: balance laws constituents}$_3$.   Then production terms $\Mi$ and $\Ei$ are obtained as corresponding moments of the collision operator,
\begin{multline}\label{production term def}
\left( \begin{matrix}
\Mi \\
E_i
\end{matrix}\right)
= \sum_{\begin{subarray}{l} j=1\\j\neq i\end{subarray}}^{I}  \int_{\mathbb{R}^3 \times \mathbb{R}_+}   \left( \begin{matrix} m_{i} \, \bv \\  \tfrac{m_{i}}{2}  \left|\bv\right|^2 + I  \end{matrix}\right) \\ \times Q_{{i j}} (M_i, M_j)(\bv,I) \, \varphi_{i} (I) \, \mathrm{d}  \bv \,\mathrm{d} I.
\end{multline}
To fulfill assumptions of Eulerian fluids \eqref{polyT: ass Eulerian} we need to specify distribution function $f_i$, that will be done in the next Section.

\section{Closure obtained from  Kinetic Theory }\label{Sec: compute}

We can close the set of equations  \eqref{polyT: balance laws constituents} obtained also by kinetic theory in the previous Section,  by means of the following  steps:
\begin{description}
	\item[i)] The Eulerian fluids \eqref{polyT: ass Eulerian} can be obtained from definitions \eqref{polyT: macro eq press and heat} by taking Maxwellian distribution function
	\begin{equation}\label{maxw distr fun}
	M_{i} 
	= \frac{n_i}{\zeta_{0_i}(T_i)} \left( \dfrac{m}{ 2 \, \pi \, k \, T_i} \right)^{3/2} e^{-\frac{1}{ k T_i} \left( \frac{m_i}{2}|v-u_i|^2+ I\right)}, 
	\end{equation}
	with macroscopic number density $n_i=\rho_i/m_i$, macroscopic velocity $\bu_i$ and temperature $T_i$ (connected to the pressure via $p_i = n_i k T_i$, $k$ being the Boltzmann constant). Namely,  plugging \eqref{maxw distr fun} into definition \eqref{polyT: macro eq press and heat}, we get that the  pressure tensor diagonalizes, with the coefficient  $n_i k T_i$ as a diagonal term, and the  heat flux vector vanishes, as in  \eqref{polyT: ass Eulerian}.\\
	
	This distribution function \eqref{maxw distr fun} corresponds to the ``mid-equilibrium'', when the  approach to equilibrium is divided  into two processes \cite{gold-sir}: i) the Maxwellization step of a species --   the approach of each distribution function to a Maxwellian distribution with its own velocity and temperature ($f_i \rightarrow M_i$),  
	and ii) the equilibration of the species, i.e. vanishing of differences in velocity and temperature among the species. 
	\item[ii)] We choose the weight function $\varphi_i(I)=I^{\alpha_i}$, with $\alpha_i > -1$ for every $i=1,\dots,I$, so that the perfect gas law for polyatomic gases can be recovered 
	\begin{equation*}
\rho_i e_i= \left( \alpha_i + \frac{5}{2} \right)\, k \, n_i \, T_i.
	\end{equation*}
	In this case, the normalization constant in \eqref{maxw distr fun} reads
	\begin{equation*}
\zeta_{0_i}(T_i) = \int_{\mathbb{R}_+}  I^{\alpha_i}  e^{- \frac{1}{k T_i} I} \, \mathrm{d} I = (k T_i)^{\alpha_i+1} \Gamma\left[\alpha_i+1\right].
	\end{equation*} 
		\item[iii)] We choose the following cross section
	\begin{equation}\label{VHS polytropic}
	\mathcal{B}_{ij}\left( \bv, \bv_*, I, I_*, r, R, \bom \right)= 4 K R^{\frac{\gamma_{ij}}{2}} \left|\bg\right|^{\gamma_{ij}} \left| \bom \cdot \hat{\bg} \right|,  
	\end{equation}
	 $\bg=\bv-\bv_*$,
	usually called the variable hard potential model with the parameter $\gamma_{ij}$ that satisfies $\gamma_{ij}=\gamma_{ji}$ and $\gamma_{ij}>-\frac{3}{4}$, and $K$ is an appropriate dimensional constant.
	The interest of this model is that it depends on one unique parameter $\gamma_{ij}$ for each couple of species,
	which can be fitted by experiments involving only macroscopic quantities.
\end{description}

It remains to compute the production terms $\Mi$ and $\Ei$ for the choices above. They will be expressed in terms of hypergeometric function given in the Appendix \ref{App}.

\subsection{Production term $\Mi$ for the momentum exchange}
 
Using the weak form \eqref{weak form}, the production term $\Mi$ from \eqref{production term def},  that corresponds to the balance law of momentum of the species $\mathcal{A}_i$ for the Euler fluids and the cross section \eqref{VHS polytropic} reads 
\begin{multline*}
\Mi =  \sum_{\begin{subarray}{l} j=1\\j\neq i\end{subarray}}^{I}  \Dint m_i \lp \bv' - \bv \rp  M_i M_{j_*}  \\
\times 4 K (1-R) R^{\frac{\gamma_{ij}+1}{2}} \inzvMvs^{ \gamma_{ij}} \\ \times \la \bom \cdot \frac{ \bv - \bv_* }{\la \bv - \bv_* \ra} \ra \md \bom \, \dsvi.
\end{multline*}
Next we change angular variable $ \bom \mapsto \bsig = \hat{\bg}  - 2 \lp \bom \cdot \hat{\bg} \rp \bom$ with Jacobian obtained in \cite{vill}
\begin{multline}\label{polyA: dsigm preko dom}
\int_{S^2} F(\bsig)\, \md \bsig = \int_{S^2} F(\mathbf{z} - 2 \lp \bom \cdot \mathbf{z}  \rp \bom) \, 2 \la \bom \cdot \mathbf{z} \ra \md \bom, 
\end{multline}
for all unit vectors $\mathbf{z}$, and for any function $F$ such that the integrals are well defined. Expressing
\beqs
\bv'-\bv = \frac{m_j}{m_i + m_j} \lp  - \bv + \bv_* + \korenij \bsig \rp,
\eeqs
the last integral becomes
\begin{multline*}
\Mi = \sjNi   \cjed \Dint   \lp  - \bv + \bv_* + \korenij \bsig \rp \\ \times e^{-\frac{1}{k T_i} (\frac{m_i}{2} \inzvui^2 + I )} e^{-\frac{1}{k T_j} (\frac{m_j}{2} \inzvsuj^2 + I_* )} \\ (1-R) R^{\frac{\gamma_{ij}+1}{2}} \inzvMvs^{ \gamma_{ij}} \md \bsig \, \dsvi,
\end{multline*}
with 
\begin{equation*}
\cjed = K  \dfrac{  n_i  n_j   \mu_{ij} (kT_i)^{-(\alpha_i+1)} (kT_j)^{-(\alpha_j+1)}}{\Gamma\left[\alpha_i+1\right]\Gamma\left[\alpha_j+1\right]}    \lp \dfrac{m_i m_j}{4 \, \pi^2 \, k^2\,  T_i T_j} \rp^{\frac{3}{2}}.
\end{equation*}

Integration with respect to $\bsig$, and then with respect to all variables except velocities $\bv$ and $\bv_*$ leads to:
\begin{multline}\label{T_i sa brzinama}
\Mi = \sjNi \cdva  \iint_{\bRN \times \bRN} \lp  - \bv + \bv_* \rp \\ \times  e^{-\frac{m_i}{2 k T_i} \inzvui^2-\frac{m_j}{2 k T_j} \inzvsuj^2 }  \inzvMvs^{\gamma_{ij}}   \md \bv \md \bv_*,
\end{multline}
where the constant is
\begin{equation*}
\cdva = \cjed  k^2 T_i T_j \frac{16 \pi}{\lp 3+\gamma_{ij} \rp \lp 5+ \gamma_{ij} \rp}.
\end{equation*}
Now we pass to the center of mass reference frame
\beq\label{smena v, v_* u relativne}
\lp \bv, \bv_* \rp \mapsto \lp \bg := \bv - \bv_*, \bG:=\frac{m_i \bv + m_j \bv_*}{m_i + m_j} \rp,
\eeq
with unit Jacobian. Then, integration  with respect to $\bG$ yields integrals which only involve the relative velocity $\bg$:
\begin{equation}\label{pomocni int}
\Mi = -  \sjNi \ctri  \int_{\bRN} \bg \, \inzg^{\gamma_{ij}}  \, e^{-a_{ij} \inzgMuiMuj^2}  \md \bg.
\end{equation}
with
\begin{equation*}
\ctri = K \dfrac{  n_i  n_j   (kT_i)^{-\alpha_i} (kT_j)^{-\alpha_j}}{\Gamma\left[\alpha_i+1\right]\Gamma\left[\alpha_j+1\right]}   \frac{16   \pi \mu_{ij}}{\lp 3+\gamma_{ij} \rp \lp 5+\gamma_{ij} \rp} \lp \frac{a_{ij}}{\pi} \rp^{\frac{3}{2}},
\end{equation*}
and 
\begin{equation}\label{aij}
a_{ij} = \lp \frac{2 \, k \, T_i}{m_i} + \frac{2 \, k \, T_j}{m_j} \rp^{-1}.
\end{equation}

In order to treat the scalar product of $\bg$ and $\bu_i - \bu_j$, we pass to  spherical coordinates for $\bg$ by taking $\widehat{\bu_i - \bu_j}$ as a zenith direction and an angle between $\bg$ and $\bu_i - \bu_j$ as an azimuthal angle $\theta$. Then, by parity arguments we have 
\begin{multline}\label{spherical for g}
 \int_{\bRN} \bg \, \inzg^{\gamma_{ij}}  \, e^{-a_{ij} \inzgMuiMuj^2}  \md \bg
 \\
 = \frac{\bu_i-\bu_j}{\left|\bu_i-\bu_j\right|} e^{-a_{ij} \inzuiMuj^2} 2 \pi \int_0^{\infty} \left|\bg\right|^{\gamma_{ij} +3} e^{-a_{ij} \left|\bg\right|^2  } \\ \times \int_{0}^{\pi} \cos\theta \ e^{2 a_{ij}  \inzuiMuj \inzg \cos\theta} \sin \theta \ \md \theta \, \md \inzg.
\end{multline}
The integral with respect to the angular variable $\theta$ can be explicitly  computed using special functions,
\begin{multline*}
\int_{0}^{\pi} \cos\theta \ e^{A_{ij} \inzg \cos\theta} \sin \theta \ \md \theta  = \int_{-1}^1 p \, e^{A_{ij} \inzg p} \, \md p \\
= \sqrt{\pi} \frac{A_{ij}}{2} \inzg \phantom{a}_0 \tilde{F}_1\left( \frac{5}{2}, \frac{A_{ij}^2}{4 } \inzg^2 \right)\\
= \frac{2}{A_{ij}^2 \inzg^2} \lp A_{ij} \inzg \cosh(A_{ij} \inzg) - \sinh(A_{ij} \inzg) \rp,
\end{multline*}
where we have denoted $A_{ij}=2 \,a_{ij} \inzuiMuj$, and the function $ \phantom{a}_0 \tilde{F}_1$ is a hypergeometric function given in \eqref{0F1}. Next, we focus on the integral
\begin{multline*}
\int_0^{\infty} \left|\bg\right|^{\gamma_{ij} +1} e^{-a_{ij}  \left|\bg\right|^2  } \\ \times \lp A_{ij} \inzg \cosh(A_{ij} \inzg) - \sinh(A_{ij} \inzg) \rp \md \inzg
\\
= \sqrt{\pi} \frac{A_{ij}}{4}  a_{ij}^{-\frac{\gamma_{ij}+5}{2}} \Gamma\left[\frac{\gamma_{ij}+5}{2}\right] \phantom{a}_1 \tilde{F}_1\left(\frac{\gamma_{ij}+5}{2}, \frac{5}{2}, \frac{A_{ij}^2}{4 a_{ij}} \right),
\end{multline*}
by means of the integral representation \eqref{hypergeometric regularized 1F1 preko 0F1}.
Therefore, we can write $\Mi$ in closed form as follows
\begin{equation}\label{polyT: Mi konacno}
\Mi = - \sjNi \cNij \Nij
\end{equation}
where the constant $\cNij$ is 
\begin{equation}\label{const Ni}
 \cNij =  K    \frac{n_i  n_j k^{-(\alpha_i+\alpha_j)}  }{ \Gamma\left[\alpha_i+1\right]\Gamma\left[\alpha_j+1\right]}  \frac{8 \pi \mu_{ij}  }{\lp \gamma_{ij} +5\rp}  \Gamma \left[  \frac{\gamma_{ij}+3}{2} \right],
\end{equation}
and 
\begin{multline*}
\Nij= \left(\bu_i - \bu_j\right)  T_i^{-\alpha_i}T_j^{-\alpha_j} \\
\times  \lp \frac{2 \, k \, T_i}{m_i} + \frac{2 \, k \, T_j}{m_j} \rp^{\frac{\gamma_{ij}}{2}}   e^{-\lp \frac{2 \, k \, T_i}{m_i} + \frac{2 \, k \, T_j}{m_j} \rp^{-1} \inzuiMuj^2} \\
\times
\phantom{a}_1 \tilde{F}_1\left(\frac{\gamma_{ij}+5}{2}, \frac{5}{2}, \lp \frac{2 \, k \, T_i}{m_i} + \frac{2 \, k \, T_j}{m_j} \rp^{-1} \inzuiMuj^2  \right),
\end{multline*}
where $\phantom{a}_1 \tilde{F}_1$ is a hypergeometric function given in \eqref{1F1}.

\subsection{Production term $\Ei$ for the energy exchange}

The full expression of the production term $\Ei$ corresponding to the energy balance law of the species $\mathcal{A}_i$ in the case of Euler fluids
and model \eqref{VHS polytropic} for the cross section, after using the weak form \eqref{weak form}, reads
\begin{multline*}
\Ei = \sjNi \Dint   \lp \frac{m_{i}}{2} \inzvp^2 + I' - \frac{m_{i} }{2} \inzv^2 - I \rp 
\\
\times M_i M_{j*}  
4 K (1-R) R^{\frac{\gamma_{ij}+1}{2}} \inzvMvs^{\gamma_{ij}} 
\\ \times \la \bom \cdot \frac{ \bv - \bv_* }{\la \bv - \bv_* \ra} \ra \md \bom \, \dsvi.
\end{multline*}
The term in the first parenthesis can be expressed in terms of non-prime variables as follows:
\begin{multline*}
\frac{m_{i} }{2}\inzvp^2 + I' - \frac{m_{i} }{2} \inzv^2 - I \\ = - \mu_{ij}  \bg \cdot \bG + \sqrt{2 \mu_{ij} R E }  \, \bg \cdot T_{\bom}\left[\hat{\bg}\right] \\ + \inzg^2 \frac{\mu_{ij}}{2} \lp 1- R\rp \lp - \frac{m_j}{ \lp m_i +m_j\rp} +  r \rp \\ + I \lp \frac{m_j}{m_i +m_j} R + r(1-R) -1 \rp
\\ + I_* \lp \frac{m_j}{m_i + m_j} R + r \lp 1- R\rp\rp,
\end{multline*}
using notation \eqref{smena v, v_* u relativne}.
We pass to the $\bsig$ notation using \eqref{polyA: dsigm preko dom}, and integration with respect to this variable yields
\begin{multline*}
\Ei = \sjNi \frac{\cjed}{\mu_{ij}} 4 \pi  \Dint  e^{-\frac{1}{k T_i} (\frac{m_i}{2} \inzvui^2 + I )} \\ \times e^{-\frac{1}{k T_j} (\frac{m_j}{2} \inzvsuj^2 + I_* )} 
 \inzg^{\gamma_{ij}} (1-R) R^{\frac{\gamma_{ij}+1}{2}} \\ \times
\left\{- \mu_{ij} \bg \cdot \bG + \inzg^2 \frac{\mu_{ij}}{2} \lp 1- R\rp \lp - \frac{m_j}{ \lp m_i +m_j\rp} +  r \rp \right. \\ \left. + I \lp \frac{m_j}{m_i +m_j} R + r(1-R) -1 \rp\right. \\ \left.+ I_* \lp \frac{m_j}{m_i + m_j} R + r \lp 1- R\rp\rp \right\} \dsvi.
\end{multline*}
Next, we pass to the reference frame of the center of mass by means of the change of variables \eqref{smena v, v_* u relativne}. Integration with respect to $\bG$,  and then with respect to $I$ and $I_*$, $r$ and $R$ yields
\begin{multline*}
\Ei = \sjNi \ctri
 \int_{\bRN} \inzg^{\gamma_{ij}}e^{-a_{ij} \inzgMuiMuj^2} \\ \times
\left\{ \lp   d_{ij} + \frac{m_i-m_j}{ \lp m_i + m_j\rp} \frac{1}{ \lp  \gamma_{ij} +7 \rp}\rp\inzg^2
+ \mathbf{e}_{ij} \cdot \bg \right.\\ \left.
+  \frac{m_j-m_i}{2\mu_{ij}\lp m_i + m_j\rp} \frac{\lp \gamma_{ij} +3 \rp}{\lp \gamma_{ij} +7 \rp} \lp k \, T_i  + k\, T_j \rp  
\right.\\ \left. + \frac{1}{2 \mu_{ij}} \lp k\, T_j - k \, T_i \rp \right\} \, \md \bg,
\end{multline*}
with $a_{ij}$ from \eqref{aij}, and 
\beq\label{solution coeff}
\begin{split}
	d_{ij} &= \mu_{ij} \lp \frac{m_i}{2 \, k \, T_i} + \frac{m_j}{2 \, k \, T_j} \rp^{-1} \lp \frac{1}{2\,k\,T_i} - \frac{1}{2 \, k \, T_j} \rp,\\
	\mathbf{e}_{ij} &= - \lp \frac{m_i}{2 \, k \, T_i} + \frac{m_j}{2 \, k \, T_j} \rp^{-1} \lp \frac{m_i}{2\,k\,T_i} \mathbf{u}_i + \frac{m_j}{2 \, k\,T_j}  \mathbf{u}_j \rp.
\end{split}
\eeq
Comparing with \eqref{pomocni int}, we recognize that the coefficient of $\mathbf{e}_{ij}$ is  $  \cNij \Nij$. Using spherical coordinates for $\bg$ as in \eqref{spherical for g}, and calculating the integral with respect to the angular variable $\theta$
\begin{equation*}
\int_{0}^{\pi} e^{A_{ij} \inzg \cos \theta} \sin \theta \, \md \theta 
= \frac{2}{A_{ij} \inzg} \sinh(A_{ij}\inzg).
\end{equation*}
with $A_{ij} = 2 a_{ij} \inzuiMuj $, we obtain
\begin{multline*}
\Ei =  \sjNi \cNij \mathbf{e}_{ij} \cdot  \Nij + \sjNi \ctri \frac{4 \pi}{A_{ij}}  e^{-a_{ij} \inzuiMuj^2} \\  \times \int_{\bRp} \inzg^{\gamma_{ij}+1}e^{-a_{ij} \inzg^2} \sinh(A_{ij}\inzg) \\ \times
\left\{   \lp d_{ij} + \frac{m_i-m_j}{ \lp m_i + m_j\rp} \frac{1}{ \lp   \gamma_{ij} +7\rp}\rp\inzg^2
\right.\\ \left.
+  \frac{m_j-m_i}{2\mu_{ij}\lp m_i + m_j\rp} \frac{\left(\gamma_{ij}+3\right)}{\lp \gamma_{ij} +7 \rp} \lp k \, T_i  + k\, T_j \rp  
\right. \\ \left. + \frac{1}{2 \mu_{ij}} \lp k\, T_j - k \, T_i \rp \right\} \md \inzg.
\end{multline*}
We then compute the integral with respect to the $\inzg$,
\begin{multline*}
 \int_{\bRp} \inzg^{\gamma_{ij}+1+\delta}e^{-a_{ij} \inzg^2} \sinh(A_{ij}\inzg) \md \inzg \\= \sqrt{\pi} \frac{A_{ij}}{4} a_{ij}^{-\frac{\gamma_{ij}+\delta +3}{2}} \Gamma\left[  \frac{\gamma_{ij}+\delta +3}{2} \right]  
 \\ \times
 \phantom{a}_1 \tilde{F}_1\left(\frac{\gamma_{ij}+\delta+3}{2}, \frac{3}{2}, \frac{A_{ij}^2}{4 a_{ij}} \right),
\end{multline*}
where $\delta=2$ or $\delta=0$, and  $\phantom{a}_1 \tilde{F}_1$ is a hypergeometric function defined in \eqref{1F1}. Substituting the coefficients, we get the final expression for the production term that corresponds to the energy balance law of the species $\mathcal{A}_i$ of Euler fluids, for the  cross section \eqref{VHS polytropic}
\begin{multline}\label{polyT: Ei konacno}
\Ei = - \sjNi \cNij \lp \frac{m_i}{2 \, k \, T_i} + \frac{m_j}{2 \, k \, T_j} \rp^{-1} \\ \times\lp \frac{m_i}{2\,k\,T_i} \mathbf{u}_i + \frac{m_j}{2 \, k\,T_j}  \mathbf{u}_j \rp \cdot  \Nij \\
+  \sjNi   \frac{2 \, \cNij }{\lp \gamma_{ij}+3 \rp } T_i^{-\alpha_i} T_j^{-\alpha_j}
\lp \frac{2 \, k \, T_i}{m_i} + \frac{2 \, k \, T_j}{m_j} \rp^{ \frac{\gamma_{ij}}{2}} \\\times e^{-\lp \frac{2 \, k \, T_i}{m_i} + \frac{2 \, k \, T_j}{m_j} \rp^{-1} \inzuiMuj^2} 
\left\{ \lp \frac{m_i}{2 \, k \, T_i} + \frac{m_j}{2 \, k \, T_j} \rp \left(\frac{\gamma_{ij}+3}{2}\right)  
\right.\\ \left.
\times
 \lp  \mu_{ij} \frac{ T_j- T_i}{m_i  T_j + m_j T_i} + \frac{m_i-m_j}{ \lp m_i + m_j\rp} \frac{1}{ \lp \gamma_{ij} +7 \rp}\rp
\right. \\ \left. \times
\phantom{a}_1\tilde{F}_1 \lp \frac{ \gamma_{ij}+5}{2} ; \frac{3}{2} ; \lp \frac{2 \, k \, T_i}{m_i} + \frac{2 \, k \, T_j}{m_j} \rp^{-1}\inzuiMuj^2 \rp
\right.\\ \left.
+ \lp   \frac{m_j-m_i}{2\mu_{ij}\lp m_i + m_j\rp} \frac{\left(\gamma_{ij}+3\right)}{\lp \gamma_{ij} +7 \rp} \lp k \, T_i  + k\, T_j \rp  
+ \frac{\lp k\, T_j - k \, T_i \rp}{2 \mu_{ij}}   \rp
\right. \\ \left. \times
\phantom{a}_1\tilde{F}_1 \lp \frac{\gamma_{ij}+3}{2} ; \frac{3}{2} ; \lp \frac{2 \, k \, T_i}{m_i} + \frac{2 \, k \, T_j}{m_j} \rp^{-1}\inzuiMuj^2 \rp  \right\},
\end{multline}
with constant $\cNij$ from \eqref{const Ni}.

\section{Comparison with the model issuing from Extended Thermodynamics in the linearized setting}\label{Sec: comparison}

The aim of this section is to compare source terms \eqref{polyT: Mi konacno} and \eqref{polyT: Ei konacno} that we have calculated using the kinetic theory of gases with the source terms \eqref{polyT: production terms ET} obtained thanks to extended thermodynamics.

Even though we restricted the attention to the description at the Euler level, this comparison 
is not possible in general, since the source terms have a completely different non-linear  structure. 
However, they can be reduced to considerably simpler form,  if we restrict the attention to processes which occur in the neighborhood of local equilibrium state determined by the average velocity $\mathbf{u}_{i} = \mathbf{u}$ and the average temperature $T_{i} = T$ of the mixture. Under this assumption, we can linearize the source terms
\begin{multline}\label{PolyT: source term app}
	A:=A(\bu_i,\bu_j,T_i,T_j) \approx A(\bu,\bu,T,T)
	\\ + \nabla A(\bu,\bu,T,T) \cdot \left[ \bu_i - \bu \quad \bu_j - \bu \quad T_i - T \quad T_j -T \right]^T,
\end{multline}
for $A$ equal to $\Mi$ or $\Ei$,which holds when $(\bu_i,\bu_j,T_i,T_j)$ is close to $(\bu,\bu,T,T)$, for any $i,j=1,\dots,I$.

Source terms \eqref{polyT: production terms ET}  issuing from extended thermodynamics can be rewritten and then approximated by
\begin{equation}\label{polyT: production terms appr ET}
\begin{split}
\Mi &\approx \sums A_{ij}(\boldsymbol{w}^0) \lp \bu_j - \bu_i \rp, \\
\Ei &\approx \bu \cdot \sums A_{ij}(\boldsymbol{w}^0) \lp \bu_j - \bu_i \rp + \sums B_{ij}(\boldsymbol{w}^0) \lp T_j - T_i \rp,
\end{split}
\end{equation}
where $\boldsymbol{w}^0:=\boldsymbol{w}(\bu,\bu,T,T)$ denotes an objective quantity  $\boldsymbol{w}(\mathbf{u}_{i}, \mathbf{u}_{j}, T_{i}, T_{j})$ evaluated at  $(\bu,\bu,T,T)$, and the coefficients are
\begin{equation*}
A_{ij}(\boldsymbol{w}^0)=
\begin{cases}
- \frac{1}{T} \alpha_{ij}(\boldsymbol{w}^0),  & j=1,\dots,I-1, \ j\neq i, \\
\frac{1}{T} \sum_{k=1}^{I-1}\alpha_{ik}(\boldsymbol{w}^0), &  j=I,
\end{cases}
\end{equation*}
\begin{equation*}
B_{ij}(\boldsymbol{w}^0)=
\begin{cases}
- \frac{1}{T^2} \beta_{ij}(\boldsymbol{w}^0),  & j=1,\dots,I-1, \ j\neq i, \\
\frac{1}{T^2} \sum_{k=1}^{I-1}\beta_{ik}(\boldsymbol{w}^0), &  j=I.
\end{cases}
\end{equation*}

Approximating source terms   \eqref{polyT: Mi konacno} and \eqref{polyT: Ei konacno} coming from kinetic theory, we obtain
\begin{equation}\label{polyT: production terms appr KT}
\begin{split}
&\Mi \approx  \sjNi \const \frac{2\mu_{ij}}{3} \lp\bu_j-\bu_i\rp\\
&\Ei \approx \bu \cdot  \sjNi \const \frac{2\mu_{ij}}{3} \lp\bu_j-\bu_i\rp \\ &+ \sjNi  \const \frac{2(\gamma_{ij}+5)}{(\gamma_{ij}+7)} \left(\frac{1}{(\gamma_{ij}+3)} + \frac{\mu_{ij}}{(m_i+m_j)}\right) k  (T_j-T_i),
\end{split}
\end{equation}
where 
\begin{equation*}
\const = \mu_{ij}^{- \frac{\gamma_{ij}}{2}}   \frac{  16K\sqrt{\pi} }{(\gamma_{ij}+5)} \frac{2^{\frac{\gamma_{ij}}{2}}n_i n_j \Gamma\left[ \frac{\gamma_{ij}+3}{2} \right]}{\Gamma\left[\alpha_i+1\right]\Gamma\left[\alpha_j+1\right]} \left(k T\right)^{-(\alpha_i+\alpha_j) +\frac{\gamma_{ij}}{2}}
\end{equation*}

Then \eqref{polyT: production terms appr ET} and \eqref{polyT: production terms appr KT} can be directly compared to obtain explicit expressions for matrices $\boldsymbol\alpha$ and $\boldsymbol\beta$ in local equilibrium. First, we obtain the off-diagonal terms:
\beqs
\begin{split}
	\aij(\boldsymbol{w}^0)&= - \const \frac{2 \mu_{ij}}{3} T,\\*
	\bij(\boldsymbol{w}^0)&= - \const \frac{2(\gamma_{ij}+5)}{(\gamma_{ij}+7)} \left(\frac{1}{(\gamma_{ij}+3)} + \frac{\mu_{ij}}{(m_i+m_j)}\right) k  T^2
\end{split}
\eeqs
for any $i=1,\dots,I-1$ and for $1 \leq j \leq I-1$ such that $j\neq i$. Next, we get the diagonal terms:
\beqs
\begin{split}
	\alpha_{ii}(\boldsymbol{w}^0)&= - \sumsl \alpha_{i\ell}(\boldsymbol{w}^0) =  \sumsl \constil \frac{2 \mu_{i\ell}}{3} T,\\*
	\beta_{ii}(\boldsymbol{w}^0) &= - \sumsl  \beta_{i\ell}(\boldsymbol{w}^0)  \\ &= \sumsl \constil \frac{2(\gamma_{i\ell}+5)}{(\gamma_{i\ell}+7)} \left(\frac{1}{(\gamma_{i\ell}+3)} + \frac{\mu_{i\ell}}{(m_i+m_\ell)}\right) k  T^2
\end{split}
\eeqs
where $i=1,\dots, I-1$. \\

We need to check positive definiteness of obtained matrices  $\boldsymbol{\alpha}$ and $\boldsymbol{\beta}$. Since their  structure is the same, we will provide the proof for $\boldsymbol{\alpha}$. Notice that the matrix $\boldsymbol{\alpha}$ is symmetric and to show its positive definiteness we use Sylvester's criterion. Firstly, the principal minor of order one is clearly positive,
\begin{equation*}
M_1 = \alpha_{11} >0.
\end{equation*}
Then, the leading principal minor of order $k$, $M_k$, with $k=2,\dots, I-1$ (for $k=I-1$ we obtain $\boldsymbol{\alpha}$ itself), can be represented as a determinant of  a sum of two matrices by separating terms on diagonal. Namely,
\begin{equation*}
M_k = \det\left( \boldsymbol{\alpha}^0_k  + \boldsymbol{\alpha}^{\mathrm{d}}_k  \right).
\end{equation*}
	 where  the elements of  $\boldsymbol{\alpha}^0_{k}$ are given with
	 \begin{equation*}
	 {\alpha}^0_{k, ij} = \begin{cases}
	 \alpha_{ij}, &\text{if} \  i\neq j, \\
	 - \sumskm \alpha_{ik}, &\text{if} \ i=j, \\
	 \end{cases}
	 \end{equation*}
	 for $i, j = 1, \dots, k$, and the diagonal matrix $\boldsymbol{\alpha}^{\mathrm{d}}$ is given by its elements $$ {\alpha}^{\mathrm{d}}_{k, ij} = - \displaystyle\sum_{\ell=k+1}^{I} \alpha_{i\ell}  \, \delta_{ij},$$ $\delta_{ij}$ being the Kronecker delta.\\ 	

Considering the determinant of  matrix $ \boldsymbol{\alpha}^0_k$, we add  all columns to the first one, or equivalently we add all rows to the first one, and obtain zeros on the first column (row), and therefore $ \boldsymbol{\alpha}^0_k $ has zero determinant for any $k$. On the other side,   $\boldsymbol{\alpha}^{\mathrm{d}}_k $ is a diagonal matrix with all positive terms, and thus its determinant is positive for any $k$. Therefore, since the determinant of two positive semi-definite matrices is greater or equal than the sum of the two corresponding determinants \cite{lin-alg}(p. 228),    we conclude that for any $k=2,\dots, I-1$ it holds
 \begin{equation*}
M_k  \geq  \det \boldsymbol{\alpha}^0_k + \det \boldsymbol{\alpha}^{\mathrm{d}}_k >0,
 \end{equation*}
which implies that $\boldsymbol{\alpha}$ is positive definite matrix.\\

Therefore, we have determined the phenomenological coefficients  $\boldsymbol{\alpha}$ and $\boldsymbol{\beta}$  of extended thermodynamics \eqref{polyT: production terms ET} for $\bu_i = \bu$ and $T_i=T$, $i=1,\dots,I$,  from the source terms provided by the kinetic theory. These are important results in further application of the multi-temperature model. For instance, when considering the shock profile solutions, as in \cite{damir1, damir2}.

\section{Acknowledgment} 
The author would like to  thank Professor Laurent Desvillettes and Professor Srboljub Simi\'c  for many fruitful discussions and inputs for this work.  This research  is supported by   the Project No. ON174016 of Ministry of Education, Science and Technological Development, Republic of Serbia.

\appendix

\section{Hypergeometric functions}\label{App}
We  introduce the regularized Kummer confluent hypergeometric function, denoted by $\phantom{.}_1 \tilde{F}_1 (a, b, z)$, with its integral representation
\begin{equation}\label{1F1}
{\Gamma\left[ b-a \right] \Gamma\left[ a \right]}  \phantom{.}_1 \tilde{F}_1 (a, b, z)   = \int_0^1 e^{z t} t^{a-1} \lp 1-t\rp^{b-a-1} \md t, 
\end{equation}
for $b>a>0,$ see \cite{abr} p. 505, relation 13.2.1. 
Next, we introduce the following function
\begin{equation}\label{0F1}
\phantom{a}_0 \tilde{F}_1 (b, z) = \frac{1}{\Gamma[b]} e^{-2 \sqrt{z}} \phantom{a}_1 F_1 (b-\frac{1}{2}, 2 b-1, 4 \sqrt{z}).
\end{equation}
The two hypergeometric functions are connected through the integral representation
\begin{equation} \label{hypergeometric regularized 1F1 preko 0F1}
\phantom{a}_1 \tilde{F}_1 (a, b, z) = \frac{1}{\Gamma\left[a\right]} \, \int_{0}^{\infty} e^{- t} \, t^{a-1} \phantom{a}_0 \tilde{F}_1 (b, z t) \mathrm{d}t,
\end{equation}
for $a>0$.

\end{document}